\documentclass[11pt]{article}

\usepackage[cp1251]{inputenc}

\newcommand{\N}{N\raise.7ex\hbox{\underline{$\circ $}}$\;$}

\begin{document}

\title{
 Classification of degenerate 4-dimensional matrices with semi-group
structure and polarization optics\\[4mm]
V.M. Red'kov, E.M. Ovsiyuk \footnote{ redkov@dragon.bas-net.by; e.ovsiyuk@mail.ru
}\\[4mm]
{\small B.I. Stepanov Institute of Physics, NAS of Belarus \\
Mozyr State Pedagogical University named after I.P. Shamyakin, Belarus\\
     }
 }

\maketitle

\begin{abstract}

In polarization optics,  an important role  play  Mueller matrices
-- real four-dimensional matrices  which describe the effect of
action of  optical elements on the polarization state  of  the
light, described by 4-dimensional Stokes vectors. An important
issue is to classify possible classes of the  Mueller matrices. In particular, of
special interest are degenerate Mueller matrices  with
vanishing  determinants.

Earlier, it was  developed a special technique of
parameterizing  arbitrary 4-dimensional matrices with the use of four 4-dimensional vector $ (k, m, l, n)$.
In the paper, a classification of   degenerate 4-dimensional real matrices of rank 1, 2, 3. is elaborated.
 To separate  possible classes of degenerate matrices of ranks 1 and 2, we
    impose  linear restrictions on  $ (k, m, l, n)$,
    which are compatible with the group multiplication law.
    All the subsets  of matrices obtained by this method,
    are either sub-groups or semigroups.
    To obtain   singular matrices of rank 3, we
      specify 16 independent possibilities
     to get the 4-dimensional matrices with zero determinant.

\end{abstract}

In polarization optics,  an important role  play  Mueller matrices
-- real four-dimensional matrices  which describe the effect of
action of  optical elements on the polarization state  of  the
light, described by 4-dimensional Stokes vectors [1]. An important
issue is to classify possible  Mueller matrices. In particular, of
special interest are degenerate Mueller matrices (matrices with
vanishing  determinant, for which the law of multiplication holds, but there
is no inverse elements). In [2], [3], it was  developed a special technique of
parameterizing  arbitrary 4-dimensional matrices on the base  of the Dirac matrices.

In particular, in the spinor basis, an arbitrary
$4 \times 4$  matrix can be  parameterized by four
4-dimensional vectors $(k,m,l,n)$

 $$
G   = \left | \begin{array}{cc}
k_{0}  +\; {\bf k} \; \vec{\sigma}   &  n_{0}  + \; {\bf n}  \; \vec{\sigma}  \\[3mm]
 l_{0}  + \; {\bf l} \; \vec{\sigma} &  m_{0} +  \; {\bf m} \; \vec{\sigma}
\end{array} \right | =
\left | \begin{array}{cc}
K  &  N \\[3mm]
 L  & M
 \end{array} \right |
. \eqno(1)
$$

The matrices $G$ will be real, if the second components of parameters are imaginary

$$
k_{2} ^{*} = -  k_{2} \;, \qquad  m_{2} ^{*} = -  m_{2}\; , \qquad
l_{2} ^{*} = -  l_{2}\; , \qquad  n_{2} ^{*} = -  n_{2} \; ,
$$
and all remaining components are real.

The law of multiplication in explicit form is

$$
k_{0}'' = k_{0}' \; k_{0} + {\bf k}' \; {\bf k} +n'_{0}\;  l_{0} +
{\bf n}' \; {\bf l} \;,
$$
$$
m_{0}'' = m_{0}' \;  m_{0} + {\bf m}' \; {\bf m}
 + l'_{0}\;  n_{0} + {\bf l}' \; {\bf n} \;,
$$
$$
n_{0}'' = k_{0}' \; n_{0} + {\bf k}' \; {\bf n}
 + n'_{0} \; m_{0} + {\bf n}' \; {\bf m} \;,
 $$
 $$
l_{0}'' = l_{0}' \;  k_{0} + {\bf l}' \; {\bf k}
 + m'_{0}  \; l_{0} + {\bf m}' \; {\bf l} \;,
$$

$$
{\bf k}'' = k'_{0} \; {\bf k} + {\bf k}' \; k_{0}  + i \;  {\bf
k}' \times {\bf k} + n_{0}' \; {\bf l} + {\bf n}'\; l_{0} + i\;
{\bf n}' \times {\bf l} \; , \qquad
$$
$$
{\bf m}'' = m'_{0} \; {\bf m} + {\bf m}' \; m_{0}  + i \;  {\bf
m}' \times {\bf m} + l_{0}' \; {\bf n} + {\bf l}'\; n_{0} + i\;
{\bf l}' \times {\bf n} \; ,
$$
$$
{\bf n}'' = k'_{0} \; {\bf n} + {\bf k}' \; n_{0}  + i \;  {\bf
k}' \times {\bf n} + n_{0}' \; {\bf m} + {\bf n}'\; m_{0} + i\;
{\bf n}' \times {\bf m} \; ,
$$
$$
{\bf l}'' = l'_{0} \; {\bf k} + {\bf l}' \; k_{0}  + i \;  {\bf
l}' \times {\bf k} + m_{0}' \; {\bf l} + {\bf m}'\; l_{0} + i\;
{\bf m}' \times {\bf l } \; . \qquad \eqno(2)
$$

In this paper,
 we will
 study degenerate 4-dimensional matrices of rank 1, 2, 3.
 To separate  possible classes of degenerate matrices of ranks 1 and 2, we
    impose  linear restrictions on  $ (k, m, l, n)$,
    which are compatible with the multiplication law (2).
    All the subsets  of matrices obtained by this method,
    are either sub-groups or semigroups.
    To obtain   singular matrices of rank 3, we
    just  specify 16 independent possibilities
     to get the 4-dimensional matrices with zero determinant.

First, consider the variants with one independent vector.

\newpage

{\bf Variant  I(k)}:
$$
{\bf n} = A  \; {\bf k} \; , \; n_{0} = \alpha \; k_{0} \; ,
$$
$$
{\bf m} = B  \; {\bf k} \; , \;  m_{0} = \beta\; k_{0} \; ,
$$
$$
{\bf l} = D  \; {\bf k} \; , \;\;    l_{0} = t \; k_{0} \; ;
\eqno(3a)
$$
there exist only 7 solutions:
$$
( K-1) \qquad
 G =\left |
\begin{array}{cc} K & 0 \\0 & 0 \end{array} \right | \; ,
$$
$$
 (K-2) \qquad
G = \left | \begin{array}{cc}
k_{0} + {\bf k} \vec{\sigma} &  0 \\
0 & k_{0} + {\bf k} \vec{\sigma}
\end{array} \right |  ,
 $$
 $$
(K-3) \qquad
G =
\left | \begin{array}{cc}
K & 0 \\
DK & 0
\end{array} \right | ,
$$
$$
 (K-4) \qquad
G =
\left | \begin{array}{cc}
K & A K  \\
0  & 0
\end{array} \right | ,
$$
$$
(K-5) \qquad
G = \left | \begin{array}{cc}
K & A K \\
D K & AD  K
\end{array}
\right | ,
$$
$$
(K-6) \qquad
G = \left | \begin{array}{cc}
k_{0}+ \vec{k} \vec{\sigma}   & Ak_{0}+ A \vec{k} \vec{\sigma} \\
tk_{0} - A^{-1} \vec{k} \vec{\sigma}  &  At k_{0}- \vec{k} \vec{\sigma}
\end{array}
\right |   ,
$$
$$
(K-7) \qquad
G = \left | \begin{array}{cc}
k_{0}+ \vec{k} \vec{\sigma}   & \alpha k_{0}+ A \vec{k} \vec{\sigma} \\
-A^{-1} ( k_{0} + \vec{k} \vec{\sigma} ) &  -A^{-1} \alpha k_{0} -  \vec{k} \vec{\sigma}
\end{array}
\right |   .
\eqno(3b)
$$

\noindent Here there are only 7 types of solutions,
among them  there is only one sub-group $(K-2)$,
the remaining 6 cases lead to the structure of the semigroup (matrices with rank 2).

\vspace{5mm}

 {\bf Variant  I(m)}
$$
{\bf n} = A  \; {\bf m} \; , \; n_{0} = \alpha \; m_{0} \; ,
$$
$$
{\bf k} = B  \; {\bf m} \; , \;  k_{0} = \beta\; m_{0} \; ,
$$
$$
{\bf l} = D  \; {\bf m} \; , \;\;  \;  l_{0} = t \; m_{0} \; ;
\eqno(4a)
$$
there exist only  7 solutions:
$$
(M-1) \qquad
G = \left | \begin{array}{cc}
0 & 0 \\
0 & M
\end{array} \right | ,
$$
$$
(M-2) \qquad
  G = \left |
\begin{array}{cc}
M & 0 \\
0 & M
\end{array} \right | ,
$$
$$
(M-3) \qquad
G = \left | \begin{array}{cc}
0 & 0 \\
D M & M
\end{array} \right | ,
$$
$$
(M-4 ) \qquad
  G = \left |
\begin{array}{cc}
0 & AM \\
0 & M
\end{array} \right |,
$$
$$
(M-5) \qquad
G = \left | \begin{array}{cc}
(A t m_{0} - \vec{m} \vec{\sigma}) &   (A  m_{0} + A \vec{m} \vec{\sigma} )\\
( t m_{0} - A^{-1} \vec{m} \vec{\sigma} ) &  ( m_{0} + \vec{m} \vec{\sigma} )
\end{array} \right | ,
$$
$$
(M-6)
\qquad
G = \left | \begin{array}{cc}
(-\alpha A^{-1}  t m_{0} - \vec{m} \vec{\sigma}) &   (\alpha   m_{0} + A \vec{m} \vec{\sigma} )\\
( A^{-1}  m_{0} + A^{-1} \vec{m} \vec{\sigma} ) &  ( m_{0} + \vec{m} \vec{\sigma} )
\end{array} \right | ,
$$
$$
(M-7) \qquad
G = \left | \begin{array}{cc}
 AD M & A M \\
D M & M
\end{array} \right |.
\eqno(4b)
$$

\noindent
All cases, except for  $(M-2)$, describe the semigroups of the rank 2.

\vspace{5mm}

{\bf Variant  I(n)}
$$
{\bf k} = A  \; {\bf n} \; , \; k_{0} = \alpha \; n_{0} \; ,
$$
$$
{\bf m} = B  \; {\bf n} \; , \;
  m_{0} = \beta\; n_{0} \; ,
  $$
  $$
  {\bf l} = D  \; {\bf n} \; ,  \; l_{0} = t \; n_{0} \; ;
\eqno(5a)
$$
there exist only  4 solutions:
$$
(N-1) \qquad
G = \left | \begin{array}{cc}
AN & N \\
0 & 0
\end{array} \right |,
$$
$$
(N-2) \qquad
G =
\left | \begin{array}{cc}
AN & N \\
A^{2} N & A N
\end{array} \right |,
$$
$$
(N-3) \qquad
G = \left | \begin{array}{cc}
\alpha  n_{0} + A{\bf n} \vec{\sigma} &    n_{0} + {\bf n} \vec{\sigma} \\[2mm]
-\alpha  A n_{0} - A^{2} {\bf n} \vec{\sigma} & -A  n_{0} - A{\bf
n} \vec{\sigma}
\end{array} \right | ,
$$
$$
(N-4) \qquad
G = \left | \begin{array}{cc}
A n_{0} + A{\bf n} \vec{\sigma} &    n_{0} + {\bf n} \vec{\sigma} \\[2mm]
\beta A n_{0} - A^{2} {\bf n} \vec{\sigma} & \beta  n_{0} - A{\bf
n} \vec{\sigma}
\end{array} \right | .
\eqno(5b)
$$
All four solutions describe the semigroup of the  rank 2.

\vspace{5mm}

{\bf Variant  I(l)}:
$$
{\bf k} = A  \; {\bf l} \; , \; k_{0} = \alpha \; l_{0} \; ,
$$
$$
{\bf m} = B  \; {\bf l} \; , \; m_{0} = \beta\; l_{0} \; ,
$$
$$
{\bf n} = D  \; {\bf l} \; ,  \; n_{0} = t \; l_{0} \; ;
\eqno(6a)
$$
there exist only  4 solutions:
$$
(L-1) \qquad
G = \left | \begin{array}{cc}
AL & 0 \\
L & 0
\end{array} \right |,
$$
$$
(L-2) \qquad
G =
\left | \begin{array}{cc}
A L & A^{2}  L \\
 L & A L
\end{array} \right |,
$$
$$
(L-3) \;\;\;
G = \left | \begin{array}{cc}
\alpha  l_{0} + A{\bf l} \vec{\sigma} &    -\alpha  A l_{0} - A^{2} {\bf l}  \\[2mm]
l_{0} + {\bf l} \vec{\sigma}
  & -A  l_{0} - A{\bf
l} \vec{\sigma}
\end{array} \right | ,
$$
$$
(L-4) \;\;\;
G = \left | \begin{array}{cc}
A l_{0} + A{\bf l} \vec{\sigma} &    \beta Al_{0} - A^{2} {\bf l} \vec{\sigma}  \\[2mm]
l_{0} + {\bf l} \vec{\sigma}
 & \beta  l_{0} - A{\bf
l} \vec{\sigma}
\end{array} \right | .
\eqno(6b)
$$
All four solutions describe the semigroup of rank 2.

We now consider the cases of two independent vectors.

\vspace{5mm}

{\bf Variant II(k, m)}
$$
{\bf n} = A  {\bf k} + B  {\bf m} \; ,\;
  n_{0} = \alpha
k_{0} + \beta  m_{0} \; ,
$$
$$
{\bf l} = C  {\bf k} + D  {\bf m} \; , \;  l_{0} = s   k_{0} +
t  m_{0} \; ;
\eqno(7a)
$$
there exist only  5 solutions:
$$
(KM-1) \qquad
G = \left | \begin{array} {cc} K &0 \\ 0 & M \end{array} \right |
,
$$
$$
(KM-2) \qquad
G = \left  | \begin{array}{cc}
K & 0 \\
D(M-K) & M
\end{array} \right | ,
$$
$$
(KM-3) \qquad
G =
\left | \begin{array}{cc}
K & B M \\
B^{-1} K & M
\end{array} \right |,
$$
$$
(KM-4) \qquad
G =
\left | \begin{array}{cc}
K & A(K-M) \\
0 & M
\end{array} \right | ,
$$
$$
(KM-5) \qquad
G =
\left | \begin{array}{cc}
K & A(K-M) \\
C(K-M) & M
\end{array} \right |.
\eqno(7b)
$$
All solutions except for the $(KM-1)$ describe the semigroup of rank 2.

\vspace{5mm}

{\bf Variant  II(l, n)}
$$
 {\bf k} = (A  {\bf l} + B  {\bf n}) \; ,
\;  k_{0} = (\alpha l_{0} + \beta  n_{0})\; ,
$$
$$
{\bf m} =( D  {\bf l}  + C  {\bf n} )  \; , \;
 m_{0} = (t  l_{0}  + s   n_{0}) \; ;
\eqno(8a)
$$
there exist only  2 solutions:
$$
(LN-1) \qquad
G =
\left | \begin{array}{cc}
A L & N \\
L & A^{-1} N
\end{array}
\right |,
$$
$$
(LN-2) \qquad
G =
\left | \begin{array}{cc}
B N & N \\
L & B^{-1} L
\end{array} \right | .
\eqno(8b)
$$
The two solutions describe the semigroup of rank 2.

\vspace{5mm}

{\bf Variant  II(k, n)}
$$
 {\bf l} = (A  {\bf k} + B  {\bf n}) \; ,
\qquad  l_{0} = (\alpha k_{0} + \beta  n_{0}) \; ,
$$
$$
{\bf m} = (D  {\bf n}  + C  {\bf k} )  \; , \qquad
 m_{0} = (t  n_{0}  + s   k_{0}) \; ;
\eqno(9a)
$$
solutions:
$$
(KN-1) \qquad
G = \left | \begin{array}{cc}
K & N \\
AK & AN
\end{array} \right |,
$$
$$
(KN-2) \qquad
G = \left | \begin{array}{cc}
K & N \\
0 & K
\end{array} \right | .
\eqno(9b)
$$
The first solution describes the group, the second -- the semigroup of rank 2.

\vspace{5mm}

{\bf Variant  II(m, l)}
$$
 {\bf n} = A  {\bf m} + B  {\bf l} \; ,
\;  l_{0} = \alpha m_{0} + \beta  l_{0} \; ,
$$
$$
{\bf k} = D  {\bf l}  + C  {\bf m}   \; , \;
 k_{0} = t  l_{0}  + s   m_{0} \; ;
\eqno(10a)
$$
there exist only  2 solutions:
$$
(ML-1) \qquad
G =  \left | \begin{array}{cc}
AL &   A M \\
L  & M
\end{array} \right |  \;  ,
$$
$$
(ML-2) \qquad
\left | \begin{array}{cc}
M & 0 \\
L & M
\end{array} \right | .
\eqno(10b)
$$
The first solution describes the group, the second -- the semigroup of rank 2.

\vspace{5mm}

Now consider the case 3 independent vectors.

{\bf Variant  I(k, m, n)}:
$$
{\bf l} = A {\bf k} + B {\bf m} + C {\bf n} \; , \qquad
l_{0} = \alpha k_{0} + \beta m_{0} + s n_{0} \; ;
\eqno(11a)
$$
there exist only  2 solutions:
$$
(KMN-1) \qquad
 G =
\left | \begin{array}{cc} K & N \\ 0 &
M \end{array} \right |  ,
$$
$$
(KMN-2) \qquad
G = \left | \begin{array}{cc} K & N \\  -K +M + N  & M \end{array}
\right |.
\eqno(11b)
$$
The first solution describes the group, the second - the semigroup of rank 2.

 There are similar solutions for  {\bf variant   I(k, m, l)}:
$$
(KML-1), \qquad
\ G  =
 \left | \begin{array}{cc} K & 0 \\ L  &
M \end{array} \right | ,
$$
$$
(KML-2), \qquad  G =
\left |
\begin{array}{cc}
K & -M +K + L \\  L  & M \end{array}
\right | .
\eqno(11c)
$$

\vspace{5mm}
{\bf  Variant  I(n, l, k)}
$$
{\bf m} = A {\bf n} + B {\bf l} + C {\bf k} \; , \qquad
m_{0} = \alpha n_{0} + \beta l_{0} + s k_{0} \; ;
\eqno(12a)
$$
there exists only  1 solution:
$$
(NLK-1) \qquad
G = \left | \begin{array}{cc}
K & N \\
L & (K  +A N -A^{-1} L)
\end{array} \right |
\eqno(12b)
$$
it describes a semigroup of rank 2.

There is a similar solution for {\bf  variant  I(n, l, m)}
$$
(NLM-1) \qquad
 G =  \left | \begin{array}{cc}
(M  +A L -A^{-1} N )  & N \\
L & M
\end{array} \right |;
\eqno(12c)
$$
it describes a semigroup of rank 2.

In all cases above, from the  semigroups of the rank 2
one can easily to obtain semi-groups of the rank 1, for this it suffices to add a requirement
that  the determinant
of the basic $2\times2$-matrices  be equal to zero.

Let us consider singular Mueller matrices of the rank 3.
Given the explicit form of the matrix $ G $, it  is easy to understand that there are 16 simple ways
to get the semigroups of rank 3. It is enough to have vanishing any $i$-line  and any $j$-column
in the original 4-dimensional matrix.
Compatibility of the law of multiplication with this constrain is obvious.

All 16 possibilities are listed below.

\vspace{2mm}

\underline{Variant $ (00) $}
$$
G
=
\left | \begin{array}{cccc}
0   &    0     & 0&   0 \\
0     & 2k_{0}        & \;\;    2n_{1}  &   2n_{0} \\
0  &     2l_{1}      & \;\;    m_{0} + m_{3} &   m_{1} - i m_{2} \\
0     & 2l_{0}        & \;\;    m_{1} + i m_{2}& m_{0} - m_{3}
\end{array} \right |,
$$
$$
k_{1} =0,\qquad  k_{2} = 0 , \qquad  k_{0} = -k_{3} \; ,
$$
$$
n_{0}= -n_{3} , \qquad l_{0}= -l_{3} , \qquad +in_{2} = n_{1} , \qquad -il_{2} = l_{1} \; .
$$

\underline{Variant  $ (01) $}
$$
G=
\left | \begin{array}{cccc}
0    &     0    & \;\;  0&  0 \\
2k_{1}     & 0        & \;\;   2 n_{1}  & 2n_{0}  \\
2l_{0}    &     0    & \;\;    m_{0} + m_{3} &   m_{1} - i m_{2} \\
2l_{1}      & 0       & \;\;    m_{1} + i m_{2}
& m_{0} - m_{3}
\end{array} \right |,
$$
$$
k_{0} =0,\qquad  k_{3} = 0 , \qquad  k_{1} = ik_{2} \; ,
$$
$$
l_{1}= il_{2} , \qquad l_{0}= l_{3} , \qquad n_{0} = -n_{3} , \qquad n_{1} = in_{2} \; .
$$

\underline{Variant  $ (02) $}
$$
G=
\left | \begin{array}{cccc}
0   &   0    & \;\;   0 &  0 \\
2k_{1}      & 2k_{0}        & \;\;   0 & 2n_{0}  \\
l_{0}+l_{3}    &     l_{1} - i l_{2}     & \;\;   0&   2m_{1}  \\
l_{1} + i l_{2}     & l_{0}-l_{3}        & \;\;    0
& 2m_{0}
\end{array} \right |,
$$
$$
n_{1} =0,\qquad  n_{2} = 0 , \qquad  n_{0} = -n_{3} \; ,
$$
$$
m_{0}= -m_{3} , \qquad m_{1}= -im_{2} , \qquad k_{0} = -k_{3} , \qquad k_{1} = ik_{2}\; .
$$

\underline{Variant  $ (03) $}
$$
G=
\left | \begin{array}{cccc}
0   &    0     & \;\;   0 &  0 \\
2k_{1}     &2 k_{0}        & \;\;    2n_{1}  & 0 \\
l_{0}+l_{3}    &     l_{1} - i l_{2}     & \;\;    2m_{0}  &   0 \\
l_{1} + i l_{2}     & l_{0}-l_{3}        & \;\;   2 m_{1}
&0
\end{array} \right |,
$$
$$
n_{0} =0,\qquad  n_{3} = 0 , \qquad  n_{1} =i n_{2} \; ,
$$
$$
m_{0}= m_{3} , \qquad m_{1}= im_{2} , \qquad k_{0} = -k_{3} , \qquad k_{1} = ik_{2} \; .
$$

\underline{Variant  $ (10) $}
$$
G=
\left | \begin{array}{cccc}
0    &     2k_{1}     & \;\;   2n_{0} &   2n_{1}  \\
0     & 0       & \;\;    0 & 0 \\
0    &     2l_{1}      & \;\;    m_{0} + m_{3} &   m_{1} - i m_{2} \\
0     & 2l_{0}        & \;\;    m_{1} + i m_{2}
& m_{0} - m_{3}
\end{array} \right |,
$$
$$
k_{0} =0,\qquad  k_{3} = 0 , \qquad  k_{1} =-i k_{2} \; ,
$$
$$
l_{0}= -l_{3} , \qquad l_{1}= -il_{2} , \qquad n_{1} = -in_{2} , \qquad n_{0} = n_{3}\; .
$$

\underline{Variant  $ (11) $}

$$
G=
\left | \begin{array}{cccc}
2k_{0}    &     0   & \;\;   2n_{0}  &  2 n_{1}  \\
0     & 0       & \;\;    0 & 0 \\
2l_{0}   &     0   & \;\;    m_{0} + m_{3} &   m_{1} - i m_{2} \\
2l_{1}     & 0        & \;\;    m_{1} + i m_{2}
& m_{0} - m_{3}
\end{array} \right | ,
$$
$$
k_{1} =0,\qquad  k_{2} = 0 , \qquad  k_{0} = k_{3} \; ,
$$
$$
l_{0}= l_{3} , \qquad l_{1}= il_{2} , \qquad n_{1} =- in_{2} , \qquad n_{0} = n_{3} \;.
$$

\underline{Variant  $ (12) $}
$$
G=
\left | \begin{array}{cccc}
2k_{0}    &     2k_{1}     & \;\;   0 &   2n_{1}  \\
0    & 0        & \;\;    0 & 0 \\
l_{0}+l_{3}    &     l_{1} - i l_{2}     & \;\;    0 &   2m_{1}  \\
l_{1} + i l_{2}     & l_{0}-l_{3}        & \;\;    0
&2 m_{0}
\end{array} \right |,
$$
$$
n_{0} =0,\qquad  n_{3} = 0 , \qquad  n_{1} =-i n_{2} \; ,
$$
$$
m_{0}= -m_{3} , \qquad m_{1}= -im_{2} , \qquad k_{1} =- ik_{2} , \qquad k_{0} = k_{3} \; .
$$

\underline{Variant   $ (13) $}
$$
G
=\left | \begin{array}{cccc}
2k_{0}    &    2 k_{1}      & \;\;   2n_{0}  &   0 \\
0     & 0        & \;\;    0 & 0\\
l_{0}+l_{3}    &     l_{1} - i l_{2}     & \;\;    2m_{0} &   0 \\
l_{1} + i l_{2}     & l_{0}-l_{3}        & \;\;    2m_{1}
& 0
\end{array} \right |,
$$
$$
n_{1} =0,\qquad  n_{2} = 0 , \qquad  n_{0} = n_{3} \; ,
$$
$$
m_{0}= m_{3} , \qquad m_{1}= im_{2} , \qquad k_{1} =- ik_{2} , \qquad k_{0} = k_{3} \; .
$$

\underline{Variant  $ (20) $}
$$
G=
\left | \begin{array}{cccc}
0    &     2k_{1}      & \;\;   n_{0} + n_{3} &   n_{1} - i n_{2} \\
0    & 2k_{0}        & \;\;    n_{1} + i n_{2} & n_{0} - n_{3} \\
0   &    0     & \;\;    0 &  0 \\
0    & 2l_{0}        & \;\;    2m_{1}
& 2m_{0}
\end{array} \right |,
$$
$$
l_{1} =0,\qquad  l_{2} = 0 , \qquad  l_{0} =- l_{3} \; ,
$$
$$
m_{0}= -m_{3} , \qquad m_{1}= im_{2} , \qquad k_{1} =- ik_{2} , \qquad k_{0} = -k_{3} \; .
$$

\underline{Variant  $ (21) $}
$$
G=
\left | \begin{array}{cccc}
2k_{0}    &    0     & \;\;   n_{0} + n_{3} &   n_{1} - i n_{2} \\
2k_{1}    & 0       & \;\;    n_{1} + i n_{2} & n_{0} - n_{3} \\
0   &     0     & \;\;    0 &   0 \\
2l_{1}      & 0        & \;\;    2m_{1}
& 2m_{0}
\end{array} \right |,
$$
$$
l_{0} =0,\qquad  l_{3} = 0 , \qquad  l_{1} =i l_{2} \; ,
$$
$$
m_{0}= -m_{3} , \qquad m_{1}= im_{2} , \qquad k_{1} = ik_{2} , \qquad k_{0} = k_{3} \; .
$$

\underline{Variant  $ (22) $}
$$
G=
\left | \begin{array}{cccc}
k_{0}+k_{3}    &     k_{1} - i k_{2}     & \;\;   0 &   2n_{1}  \\
k_{1} + i k_{2}     & k_{0}-k_{3}        & \;\;   0 & 2n_{0}  \\
0   &   0     & \;\;    0 &   0 \\
2l_{1}      & 2l_{0}      & \;\;    0
& 2m_{0}
\end{array} \right |,
$$
$$
m_{1} =0,\qquad  m_{2} = 0 , \qquad  m_{0} =-m_{3} \; ,
$$
$$
n_{0}= -n_{3} , \qquad n_{1}=- in_{2} , \qquad l_{1} = il_{2} , \qquad l_{0} = -l_{3} \; .
$$

\underline{Variant  $ (23) $}
$$
G=
\left | \begin{array}{cccc}
k_{0}+k_{3}    &     k_{1} - i k_{2}     & \;\;  2 n_{0}&  0 \\
k_{1} + i k_{2}     & k_{0}-k_{3}        & \;\;    2n_{1}  & 0\\
0   &    0     & \;\;   0&   0 \\
2l_{1}      & 2l_{0}       & \;\;    2m_{1}
&0
\end{array} \right |
$$
$$
m_{0} =0,\qquad  m_{3} = 0 , \qquad  m_{1} =im_{2} \; ,
$$
$$
n_{0}= n_{3} , \qquad n_{1}= in_{2} , \qquad l_{1} = il_{2} , \qquad l_{0} = -l_{3}\; .
$$

 {\bf Variant $(30)$}
$$
G=
\left | \begin{array}{cccc}
0    &     2k_{1}      & \;\;   n_{0} + n_{3} &   n_{1} - i n_{2} \\
0    &2 k_{0}        & \;\;    n_{1} + i n_{2} & n_{0} - n_{3} \\
0   &     2l_{1}      & \;\;   2 m_{0}  &   2m_{1} \\
0     & 0      & \;\;    0
& 0
\end{array} \right |,
$$
$$
l_{0} =0,\qquad  l_{3} = 0 , \qquad  l_{1} =-il_{2} \; ,
$$
$$
k_{0}= -k_{3} , \qquad k_{1}= -ik_{2} , \qquad m_{1} = -im_{2} , \qquad m_{0} = m_{3}\; .
$$

\underline{Variant  $ (31) $}
$$
G=
\left | \begin{array}{cccc}
2k_{0}    &   0   & \;\;   n_{0} + n_{3} &   n_{1} - i n_{2} \\
2k_{1}    & 0        & \;\;    n_{1} + i n_{2} & n_{0} - n_{3} \\
2l_{0}   &     0    & \;\;    2m_{0}  &   2m_{1}  \\
0    & 0        & \;\;    0
& 0
\end{array} \right |,
$$
$$
l_{1} =0,\qquad  l_{2} = 0 , \qquad  l_{0} =l_{3} \; ,
$$
$$
k_{0}= k_{3} , \qquad k_{1}= ik_{2} , \qquad m_{1} = -im_{2} , \qquad m_{0} = m_{3} \; .
$$

\underline{Variant  $ (32) $}
$$
G=
\left | \begin{array}{cccc}
k_{0}+k_{3}    &     k_{1} - i k_{2}     & \;\;   0 &  2 n_{1} \\
k_{1} + i k_{2}     & k_{0}-k_{3}        & \;\;    0& 2n_{0}  \\
2l_{0}    &    2 l_{1}    & \;\;   0 &   2m_{1}  \\
0     & 0        & \;\;    0
& 0
\end{array} \right |,
$$
$$
m_{0} =0,\qquad m_{3} = 0 , \qquad  m_{1} =-im_{2} \; ,
$$
$$
l_{0}= l_{3} , \qquad l_{1}= -il_{2} , \qquad n_{1} = -in_{2} , \qquad n_{0} =- n_{3}\; .
\eqno(14c)
$$

\underline{Variant  $ (33) $}
$$
G=
\left | \begin{array}{cccc}
k_{0}+k_{3}    &     k_{1} - i k_{2}     & \;\;   2n_{0}  &  0 \\
k_{1} + i k_{2}     & k_{0}-k_{3}        & \;\;    2n_{1} & 0 \\
2l_{0}    &     2l_{1}     & \;\;   2 m_{0} &   0 \\
0     & 0       & \;\;    0
& 0
\end{array} \right |,
$$
$$
m_{1} =0,\qquad m_{2} = 0 , \qquad  m_{0} =m_{3} \; ,
$$
$$
l_{0}= l_{3} , \qquad l_{1}= -il_{2} , \qquad n_{1} = in_{2} , \qquad n_{0} = n_{3} \; .
$$

\vspace{3mm}

REFERENCES

\vspace{3mm}

    1. Snopko, V.N. Polarization characteristics of optical radiation and their measurement methods
 / V.N. Snopko. -- Minsk: Nauka i tehnika, 1992. -- 334 p.

2. Bogush, A.A. About four-vector parametrization of the group and some of its subgroups
 / A.A. Bogush, V.M. Red'kov // Proc. of the Natl. Academy of Sciences of Belarus, Ser. Phys.-Math. Sci. --
 2006. -- no 3. -- P. 57--63.

3. Bogush, A.A. On Unique parametrization of the linear group
GL(4.C) and its subgroups by using the Dirac algebra basis / A.A.
Bogush, V.M. Red'kov //
 NPCS. -- 2008. -- Vol. 11, no 1. -- P. 1--24.

\end{document}